\documentclass[conference]{IEEEtran}
\IEEEoverridecommandlockouts
\usepackage{cite}
\usepackage{amsmath,amssymb,amsfonts}
\usepackage{listings} 
\usepackage{algorithmic}
\usepackage{graphicx}
\usepackage{textcomp}
\usepackage{xcolor}
\usepackage{tabularray}
\usepackage[colorlinks=true,linkcolor=black,anchorcolor=black,citecolor=black,filecolor=black,menucolor=black,runcolor=black,urlcolor=black]{hyperref}
\usepackage{color}
\usepackage{array}
\usepackage{multirow}
\usepackage{ragged2e}
\usepackage{scalerel}
\usepackage{tikz}
\usetikzlibrary{svg.path}

\definecolor{orcidlogocol}{HTML}{A6CE39}
\tikzset{
  orcidlogo/.pic={
    \fill[orcidlogocol] svg{M256,128c0,70.7-57.3,128-128,128C57.3,256,0,198.7,0,128C0,57.3,57.3,0,128,0C198.7,0,256,57.3,256,128z};
    \fill[white] svg{M86.3,186.2H70.9V79.1h15.4v48.4V186.2z}
                 svg{M108.9,79.1h41.6c39.6,0,57,28.3,57,53.6c0,27.5-21.5,53.6-56.8,53.6h-41.8V79.1z M124.3,172.4h24.5c34.9,0,42.9-26.5,42.9-39.7c0-21.5-13.7-39.7-43.7-39.7h-23.7V172.4z}
                 svg{M88.7,56.8c0,5.5-4.5,10.1-10.1,10.1c-5.6,0-10.1-4.6-10.1-10.1c0-5.6,4.5-10.1,10.1-10.1C84.2,46.7,88.7,51.3,88.7,56.8z};
  }
}

\newcommand\orcidicon[1]{\href{https://orcid.org/#1}{\mbox{\scalerel*{
\begin{tikzpicture}[yscale=-1,transform shape]
\pic{orcidlogo};
\end{tikzpicture}
}{|}}}}

\usepackage{hyperref} 

\def\BibTeX{{\rm B\kern-.05em{\sc i\kern-.025em b}\kern-.08em
    T\kern-.1667em\lower.7ex\hbox{E}\kern-.125emX}}

    \makeatletter
    \newcommand{\linebreakand}{%
      \end{@IEEEauthorhalign}
      \hfill\mbox{}\par
      \mbox{}\hfill\begin{@IEEEauthorhalign}
    }
    \makeatother

\makeatletter

\def\ps@headings{%
\def\@oddhead{\parbox[t][\height][t]{\textwidth}{\centering
The 6th Iranian International Conference on Microelectronics (IICM2024)\\
}}%
\def\@evenhead{\parbox[t][\height][t]{\textwidth}{\centering
The 6th Iranian International Conference on Microelectronics (IICM2024)\\
}}%
}

\def\ps@IEEEtitlepagestyle{%
\def\@oddhead{\parbox[t][\height][t]{\textwidth}{\centering
The 6th Iranian International Conference on Microelectronics (IICM2024)\\
}}%
}

\makeatother
\begin{document}

\title{Evaluation of Run-Time Energy Efficiency using Controlled Approximation in a RISC-V Core}

\author{
\centering
\IEEEauthorblockN{A. Delavari~\textsuperscript{\orcidicon{0009-0006-6350-0055}}\, F. Ghoreishy~\textsuperscript{\orcidicon{0009-0007-7007-3386}}\, H. S. Shahhoseini~\textsuperscript{\orcidicon{0000-0002-6042-0993}}\, S. Mirzakuchaki~\textsuperscript{\orcidicon{0000-0003-0232-9267}}}
\IEEEauthorblockA{\textit{School of Electrical Engineering} \\
\textit{Iran University of Science and Technology}\\
Tehran, Iran \\
\{arvin\_delavari, faraz\_ghoreishy\}@elec.iust.ac.ir, \{shahhoseini, m\_kuchaki\}@iust.ac.ir}
}

\maketitle

\begin{abstract}
    The limited energy available in most embedded systems poses a significant challenge in enhancing the performance of embedded processors and microcontrollers. One promising approach to address this challenge is the use of approximate computing, which can be implemented in both hardware and software layers to balance the trade-off between performance and power consumption. In this study, the impact of dynamic hardware approximation methods on the run-time energy efficiency of a RISC-V embedded processor with specialized features for approximate computing is investigated. The results indicate that the platform achieves an average energy efficiency of 13.3 pJ/instruction at a 500MHz clock frequency adhering approximation in 45nm CMOS technology. Compared to accurate circuits and computation, the approximate computing techniques in the processing core resulted in a significant improvement of 9.21\% in overall energy efficiency, 60.83\% in multiplication instructions, 14.64\% in execution stage, and 9.23\% in overall power consumption. 
\end{abstract}


\section{Introduction}\label{introduction-section}

    The increasing demand for energy-efficient and low-power embedded systems has driven research efforts towards enhancing the overall performance of embedded processors, especially microcontrollers. This is an important challenge, as the limited energy available in most embedded applications imposes constraints on the design and operation of these devices. A promising approach to address this challenge is the use of approximate computing, which can be implemented in both hardware and software layers to balance the trade-off between performance and power consumption. Approximation is allowed in error-resilient applications such as image processing \cite{ref-1}, neural networks \cite{ref-2}, Processing in Memory (PIM) \cite{ref-3} and etc.
  
    The RISC-V instruction set architecture has emerged as a popular choice for embedded processor designers, offering a flexible platform for a wide range of applications \cite{ref-4} \cite{ref-5}. The integration of RISC-V and approximate computing techniques presents an opportunity to further improve the energy efficiency and performance of embedded processors.

    This study investigates the impact of dynamic hardware approximation methods on the run-time energy efficiency of a RISC-V embedded processor with specialized features for approximate computing. The results demonstrate that the platform achieves a significant improvement in energy efficiency, area, and power consumption compared to accurate circuits and computation, highlighting the potential of this approach for enhancing the performance of embedded systems. Main contributions of this paper are as follows:

    \begin{itemize}
    
        \item Utilizing a RISC-V embedded processor platform Hardware/Software interface method using standard RISC-V CSRs and special programming conventions for approximate computing.
        
        \item Design of error-configurable adder/subtracter and multiplier circuit for the experimental setup.
        
        \item Analysis and evaluation of approximation and fault-injection methods efficiency at core level, in an embedded processor.
    
    \end{itemize}

    In the remainder of this paper, we will present: a review of related previous works in this subject (Section \ref{related-works-section}), a description of the experimental setup used in this research (Section \ref{execution-engine-section}), the results of the experiments and a comparison of the findings (Section \ref{results-section}), and an overall summary of the project along with concluding remarks (Section \ref{conclusion-section}).

\section{Related Works}\label{related-works-section}
    
    Approximate computing has enhanced energy efficiency in many applications in recent years. There has been exploration into combining RISC-V with approximate computing methods. Software adaptations have proposed ISA extensions \cite{ref-6} and multi-level accuracy control mechanisms to enable this. The AxPIKE ISA simulator \cite{ref-7} enables incorporating hardware approximation at the instruction level to assess quality-energy trade-offs. Other studies have characterized RISC-V architecture extensions that orchestrate diverse circuit-level approximation techniques, especially approximate multipliers \cite{ref-8}. Customized RISC-V cores have also integrated approximate multipliers to demonstrate energy-efficient execution of neural networks \cite{ref-9}.
   
    There have been investigations on the advantages of multi-core systems that combine exact and approximate compute cores \cite{ref-10}. This heterogeneous approach aims to leverage the benefits of approximation while maintaining the precision of exact computation. \cite{ref-11} proposes an approach to improve the performance of multi-core systems by configurable approximate arithmetic logic units, with a machine learning based framework for online power regulation and quality monitoring to adjust the frequency and precision of the arithmetic units and maximize performance.

    The proposed design here involves a RISC-V core where each execution unit can dynamically switch between accurate and approximate circuits and also if applicable control the error level of integrated approximate execution units. This combines the efficiency gains of dynamic approximation with the precision of exact computation all within a single reconfigurable core.

\section{Experimental Setup}\label{execution-engine-section}

\begin{figure}[htbp]
    \centering
    \includegraphics[width=0.9\linewidth]{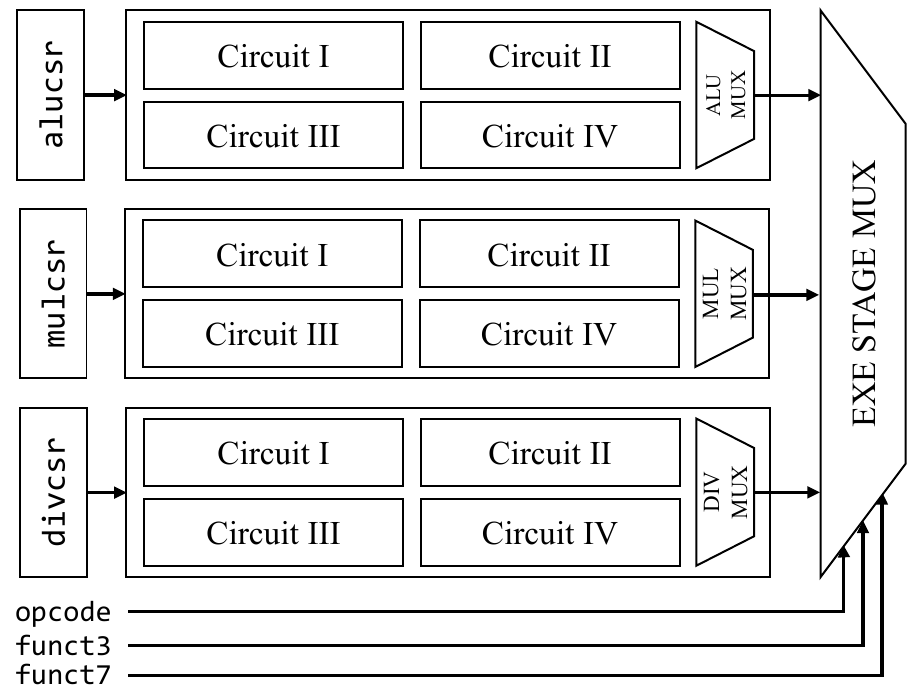}
    \caption{Execution stage high-level block diagram of the processor including accurate and approximate arithmetic circuits}
    \label{fig-1}
    \end{figure}

    The considered platform for these experiments is an embedded processor named phoeniX \cite{ref-12} which supports RV32IEM standard instruction set of the open-source and license-free RISC-V ISA. The chosen core is a 3-stage pipelined scalar processor with specialized and novel features for approximate computing, in order to be a general platform for fault-tolerant applications and approximation-enhanced low-power design studies. In this study, “fault-tolerant” refers to applications that can accommodate errors (specifically approximations) resulting from their functions and data. Pipeline stages are considered as Instruction Fetch (IF) and Decode (ID), Execution (EXE) and Memory access with Write-back (MEM/WB). The high-level architecture of the processor’s execution stage within the pipeline is illustrated in Fig.~\ref{fig-1}.

    Execution stage consists of Arithmetic Logic Unit, Multiplier Unit and Divider Unit due to the support of RV32IEM instruction set. These three computational units share a unique method for adhering and controlling approximation in circuit-level. The approximation level is determined from a customized special-purpose control status registers addressed in \textit{0x800}, \textit{0x801} and \textit{0x802} in standard RISC-V CSR addressing space \cite{ref-5}. Each execution unit can host four arithmetic circuit (accurate or approximate) supported by the standard signaling conventions of the processor’s execution engine. The execution unit’s active circuit and target result along with approximation control logic for each is determined by the value in the CSRs, which the decoding and framing is explained further in this paper.
    
    These CSRs are named \verb|alucsr|, \verb|mulcsr| and \verb|divcsr|, designed for enabling and controlling approximation level of the dynamically configurable circuits. Each field of CSRs is responsible for a feature in approximation control and advancement. Bit 0 is for enabling approximation, in which approximate arithmetic is used when signal value is 1, and will perform accurate arithmetic with value of 0. Bits 2 to 1 are for determination of selected circuit, for controlling result multiplexer and switching off unused circuits. Bits 3 to 7 are used for dynamic truncation control in arithmetic circuits which may support this technique. Bits 8 to 11 and bits 12 to 15 are also custom fields of control status registers for designer-defined control logic and features. In the end, bits 16 to 31 are responsible for dynamic error control in case of integration of error controllable arithmetic circuits within the core.

\begin{figure}[htbp]
    \centering
    \includegraphics[width=0.90\linewidth]{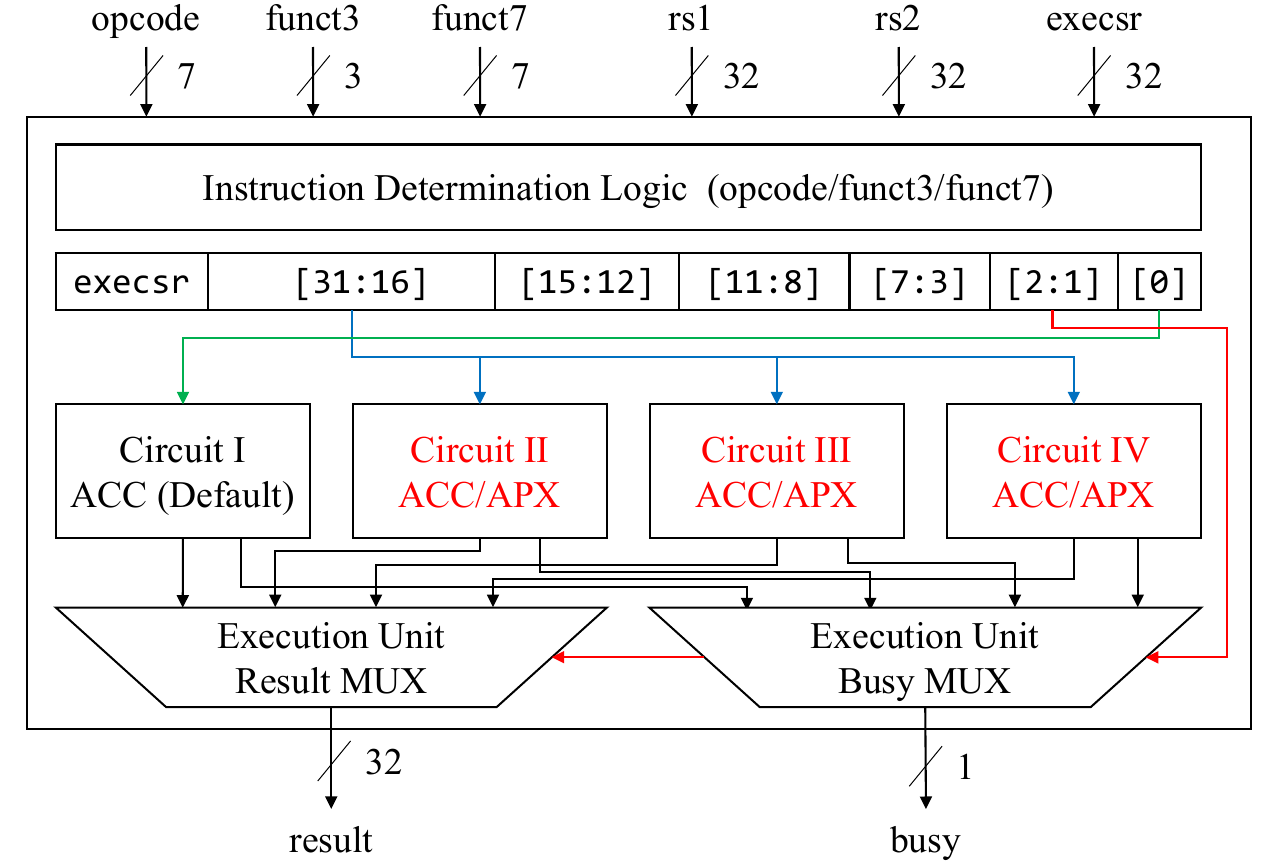}
    \caption{ Error control, circuit switching, and hardware approximation level determination based on custom CSR values in EXE stage of processor}
    \label{fig-2}
    \end{figure}

    Each execution unit can embed up to 4 arithmetic circuit with platform’s special conventions for circuit integration. Circuits can be accurate, approximate with static configurations and error, or dynamic error-configurable such as selected circuits in this study. The platform can adhere and synchronize execution process with different circuit which may vary in timing and other hardware specifications. Block diagram of the execution units’ structure is shown in Fig.~\ref{fig-2} In this experiment an error-configurable multiplier, alongside an accurate multiplier, and fully accurate division circuit is considered, because of the importance of multiplication and accumulation in most computational workloads, including fault-tolerant applications. The accurate multiplier is positioned in the default slot (Circuit I), while the approximate multiplier is placed in the secondary slot (Circuit II). The remaining slots are unutilized and remained empty in this configuration.
    
    After determination of the operating circuit through the respected CSRs of each execution unit, the other circuits will be switched off in order to exclude redundant processing of data and unnecessary static and dynamic power consumption. It is important to note that all execution units including adders in ALU Arithmetic Logic Unit, Multiplier Unit and Divider unit are capable of approximation features explained, except for addition required in address generation and other control flow instructions which uses an accurate adder because approximation is not allowed in such operations.

    \subsection{Dynamic Accuracy-Controllable Carry Select Adder}
\begin{figure}[htbp]
    \centering
    \includegraphics[width=\linewidth]{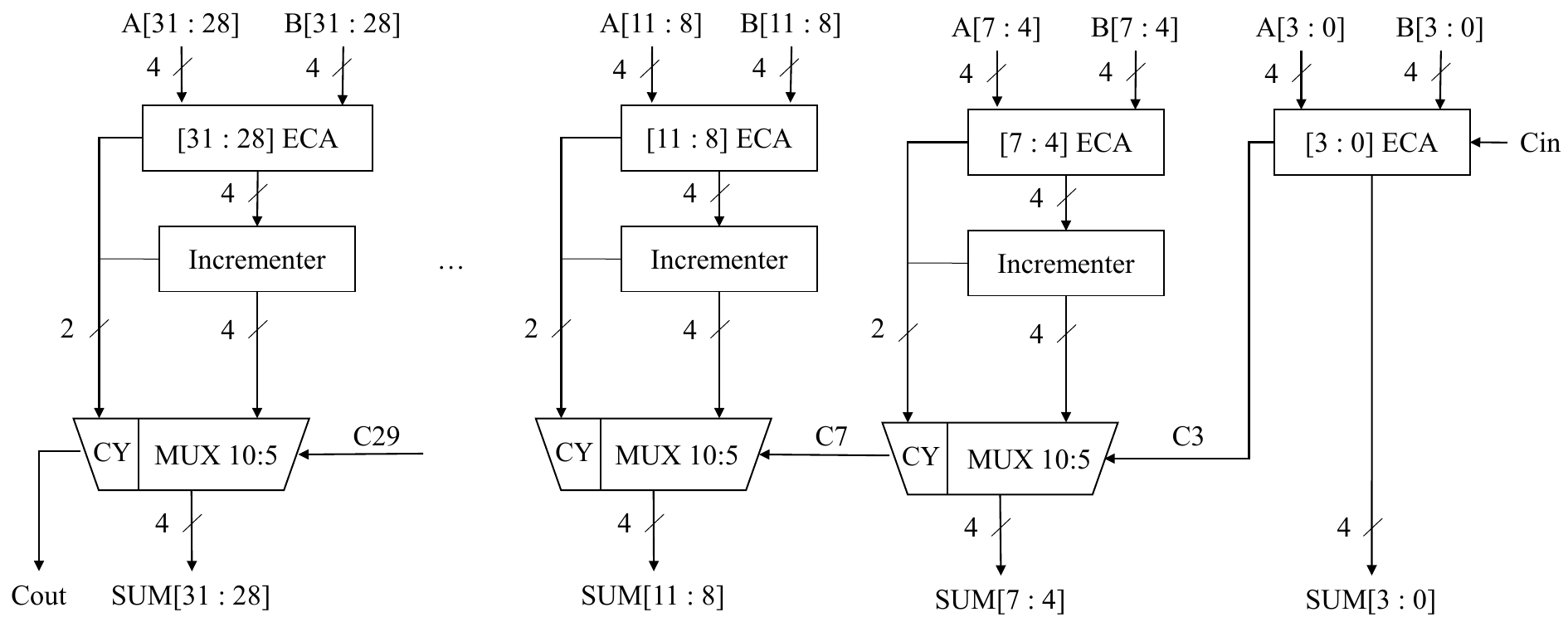}
    \caption{Error controllable low-power and fast  carry select adder}
    \label{fig-3}
    \end{figure}

    The adder/subtracter integrated within the processor, is a fast and low power carry select adder, with improved performance by decreasing the critical path in comparison with conventional design. Also, the proposed adder, depicted in Fig.~\ref{fig-3}, is an error-controllable design which can be used in different applications due to controllable accuracy of the results. The adder inside arithmetic logic unit is responsible for addition and subtraction of I/E extension of RISC-V instruction set architecture. In the experiments, the adder was set to default mode which enables accurate addition. In fact, the proposed adder can perform both accurate and approximate addition in a single circuit, without any need of additional hardware for circuit switching in the execution.

\begin{figure}[htbp]
    \centering
    \includegraphics[width=0.65\linewidth]{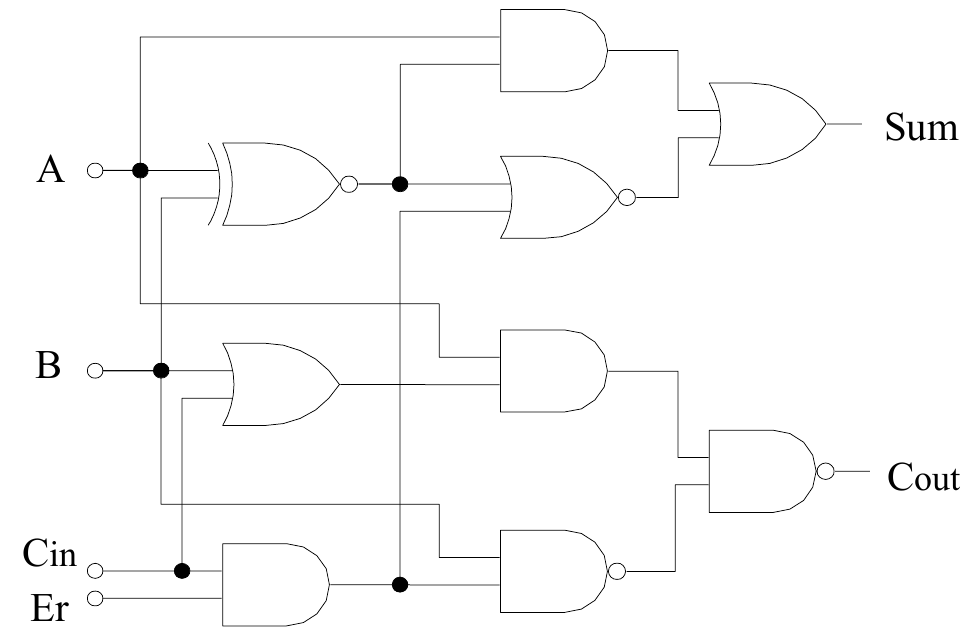}
    \caption{Proposed full-adder with error signal used in 4-bit ripple carry adders of the CSA circuit}
    \label{fig-4}
    \end{figure}

   The ECA shown in Fig.~\ref{fig-3} is a 4-bit error controllable ripple carry adder, in which the fundamental full-adders have an additional control signal in comparison with conventional full adder. This full adder as shown in Fig.~\ref{fig-4}, can perform accurately when the error control signal is high (logical value 1), but may produce inaccurate result by decreasing or increasing carry value when the signal is low. When using approximation, the error in the final result is applied unevenly, as in half the input scenarios, the approximate result is lower than the accurate result by one, and in the other half, the approximate result is higher than the accurate result by one.
    
    \subsection{Dynamic Accuracy-Controllable Array Multiplier}
\begin{figure}[htbp]
    \centering
    \includegraphics[width=0.60\linewidth]{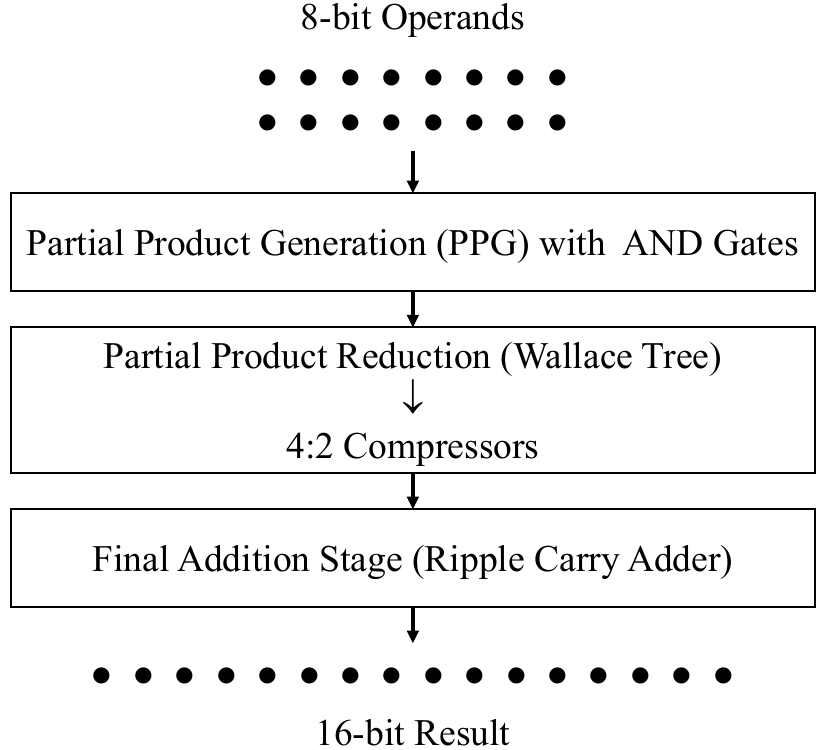}
    \caption{8-bit multiplier's high level block diagram}
    \label{fig-5}
    \end{figure}

    For this setup, an original low-power approximate 8x8 unsigned multiplier with reduced circuital path and area in comparison with a conventional Wallace tree multiplier is designed. The design also incorporates the full-adder presented in Fig.~\ref{fig-4}. The unbiased error of the adder (having both positive and negative error in results, in error enabled modes) improves the multiplier with Error Rate (ER) and Mean Relative Error Distance (MRED) \cite{ref-13} \cite{ref-14} in term of accuracy. The full-adder illustrated in Fig.~\ref{fig-4}, is integrated in the final addition stage, after partial product generation and compression stages of Wallace multiplier which is shown in Fig.~\ref{fig-5}.
    
    The ripple carry adder’s error in the final stage can be controlled using the error control signals of the full-adders. 7-bits of the 12-bit addition is controllable, which gives 128 configurations and error levels in one circuit, which means 128 configurations of dynamic power consumption. The mode with all error control signals equal to zero is the configuration with lowest power consumption, which also has the highest application-level quality in error (ER and MRED).
.

\begin{figure}[htbp]
    \centering
    \includegraphics[width=\linewidth]{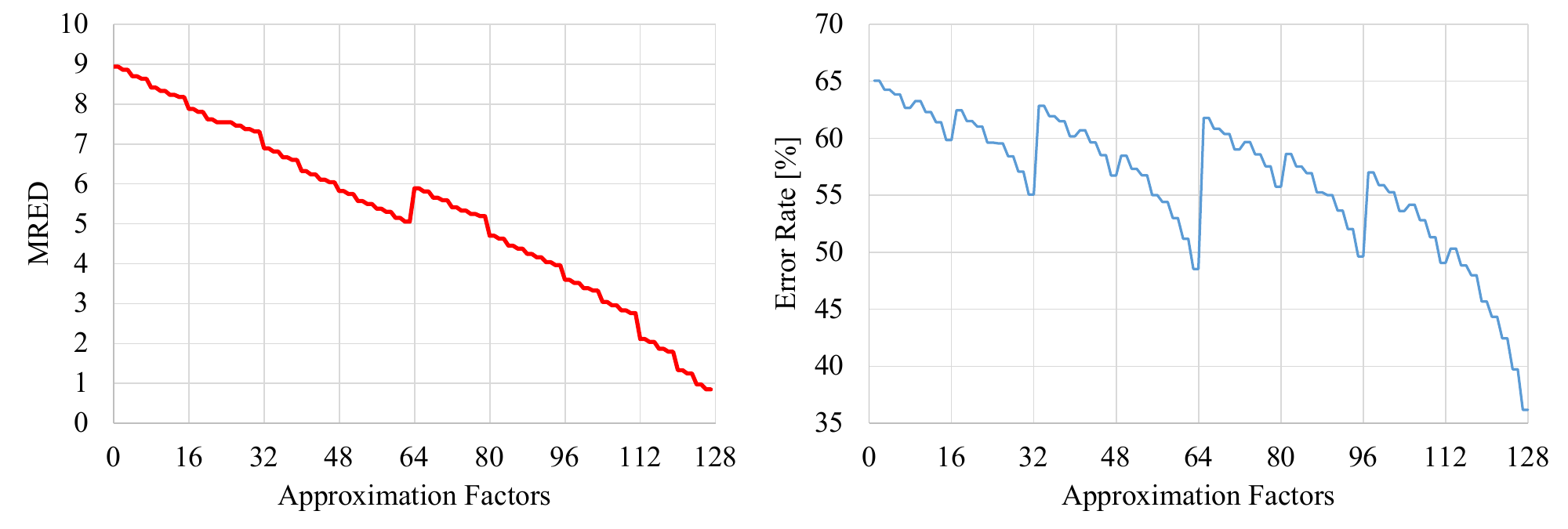}
    \caption{Error Rate and Mean Relative Error Distance (MRED) analysis of the proposed multiplier in all dynamic configurations (128 Approximation Factors)}
    \label{fig-6}
    \end{figure}

    The proposed Multiplier has an area of 269µm² with a range of power between 70.2µW and 101.3µW in different configurations. Fig.~\ref{fig-6} shows ER and MRED for all configurations of the proposed design.
    
    Table~\ref{table-1} shows an analysis in terms of hardware (power, delay and area) and accuracy efficiency. The proposed design showcases the best result in terms of critical path delay and area occupation. Additionally, in one of the configurations the design has a power consumption of 70.2µW which is the lowest in the comparison. The design in \cite{ref-10} produces the best results in accuracy efficiency criteria. 

    \begin{table}[htbp]
    \caption{Hardware - Accuracy Efficiency Comparison of Proposed Multiplier and Similar Approximate Multipliers}
    \begin{center}

    \noindent\begin{minipage}{\linewidth}
    \centering
    \begin{tblr}{
      width = \linewidth,
      colspec = {Q[120]Q[100]Q[100]Q[100]Q[85]Q[100]},
      cells = {c},
      hlines,
      vlines,
    }
    Multipliers & Power (µW) & Delay (ns) & Area (µm²) & ER 
    
    (\%) & MRED (\%)\\
    Proposed Design & \textbf{70.2} - 101.3 & \textbf{0.64} & \textbf{269.6} & 36.16 - 65.06 & 0.85 - 8.94 \\
    \cite{ref-15} & 81.2 & \textbf{0.64} & 27.8.8 & \textbf{32.42} & \textbf{0.18} \\
    \cite{ref-16} & 156.7 & 0.80 & 323.9 & 48.35 & 1.26\\
    \cite{ref-16} & 143.5 & 0.77 & 301.6 & 69.73 & 2.83
    
    \end{tblr}
    \end{minipage}
    
    \label{table-1}
    \end{center}
    \end{table}

    \vspace{-3mm}
    
    \subsection{Hierarchical 32×32 Approximate Multiplier Design} 
    For final integration of multiplier within the core, a 32-bit multiplier architecture consisting of several 8-bit modules is designed. The method used for creation of the 32-bit architecture is a hierarchical configuration of 8-bit multipliers. The original 8-bit multiplier is used over multiple cycles to perform 16-bit multiplication. This hardware is replicated four times to perform a 32-bit multiplication. There is also an accurate multiplier integrated within the core in order to have a comprehensive comparison in term of power and energy efficiency gains, using dynamic circuit switching feature which is presented in the paper. The accurate multiplier circuit is an original 32-bit multiplier implemented with DesignWare software, offered by Synopsys Design Compiler.

\section{Results and Comparison}\label{results-section}
    
    The processor and it’s approximate and accurate execution units are implemented in 45nm CMOS technology using Design Compiler, enabling a clock frequency of 500MHz which is considered for this set of experiments. All power and energy consumption data reported in this paper are extracted with post-synthesis simulation based on switching activity files.

\begin{figure}[htbp]
    \centering
    \includegraphics[width=0.95\linewidth]{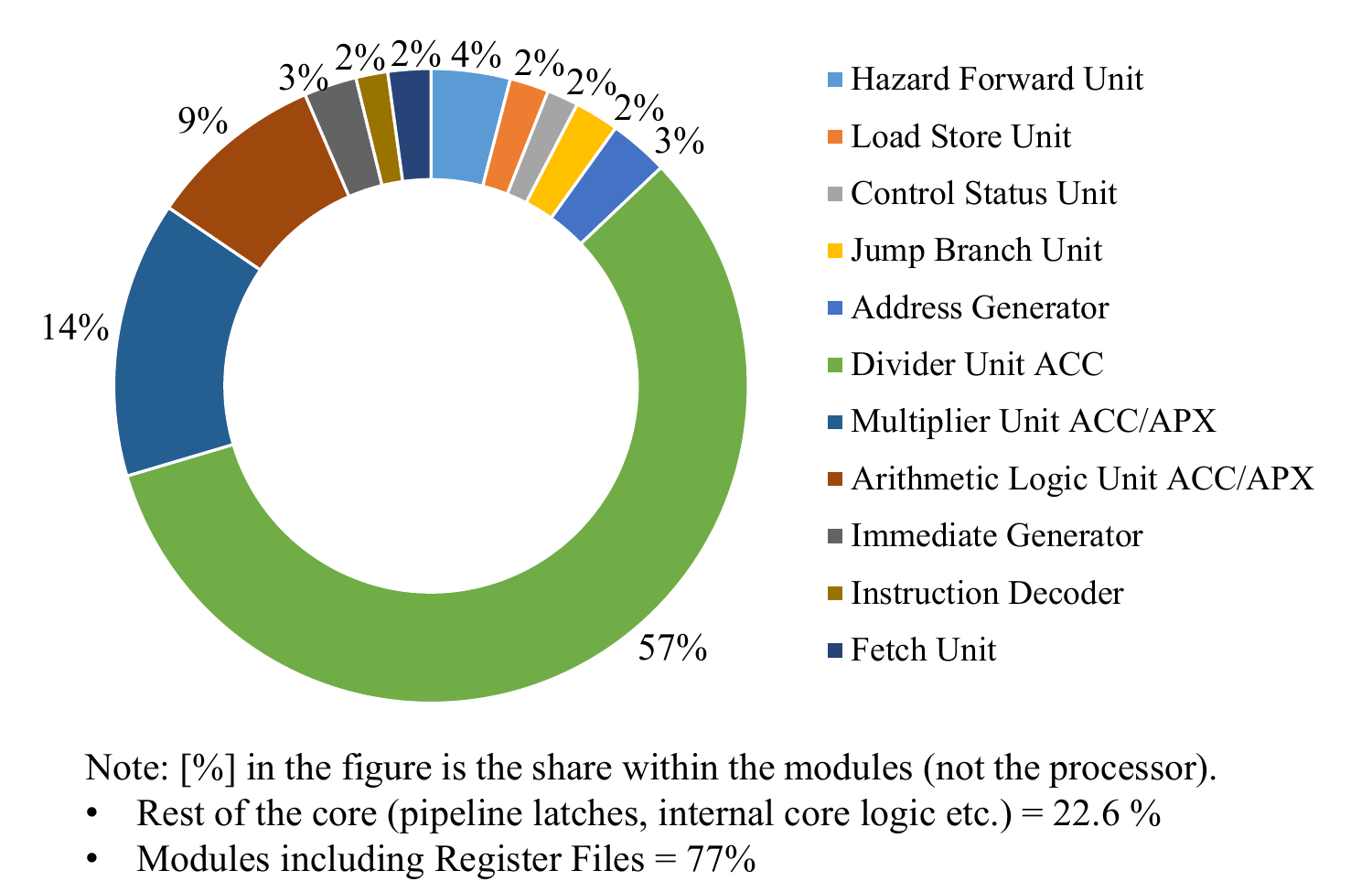}
    \caption{Modules power distribution in the processor's inner modules excluding memories and register files}
    \label{fig-7}
    \end{figure}

    Before delving into the main experiment and comparisons, Fig.~\ref{fig-7} is shown for an estimation of the power distribution in the processor, excluding memories and register files.
    
   It can be concluded from Fig.~\ref{fig-7} that the execution stage has the highest power consumption in the inner core microarchitecture. In the target processor, the execution stage, including the arithmetic logic unit, multiplier unit (with 2 multipliers), and divider unit, have the largest share of the power distribution chart, accounting for 80\%. The results provide motivation for using approximation techniques not just in software or memory approaches, but also at the circuit level. The execution stage, excluding memories and register files, consumes the most power within the inner modules. This is the reason for the focus of this study, the impact of user-controlled circuit level approximation on the overall performance of a processor.

    \vspace{-2mm}
   
\begin{figure}[htbp]
    \centering
    \includegraphics[width=\linewidth]{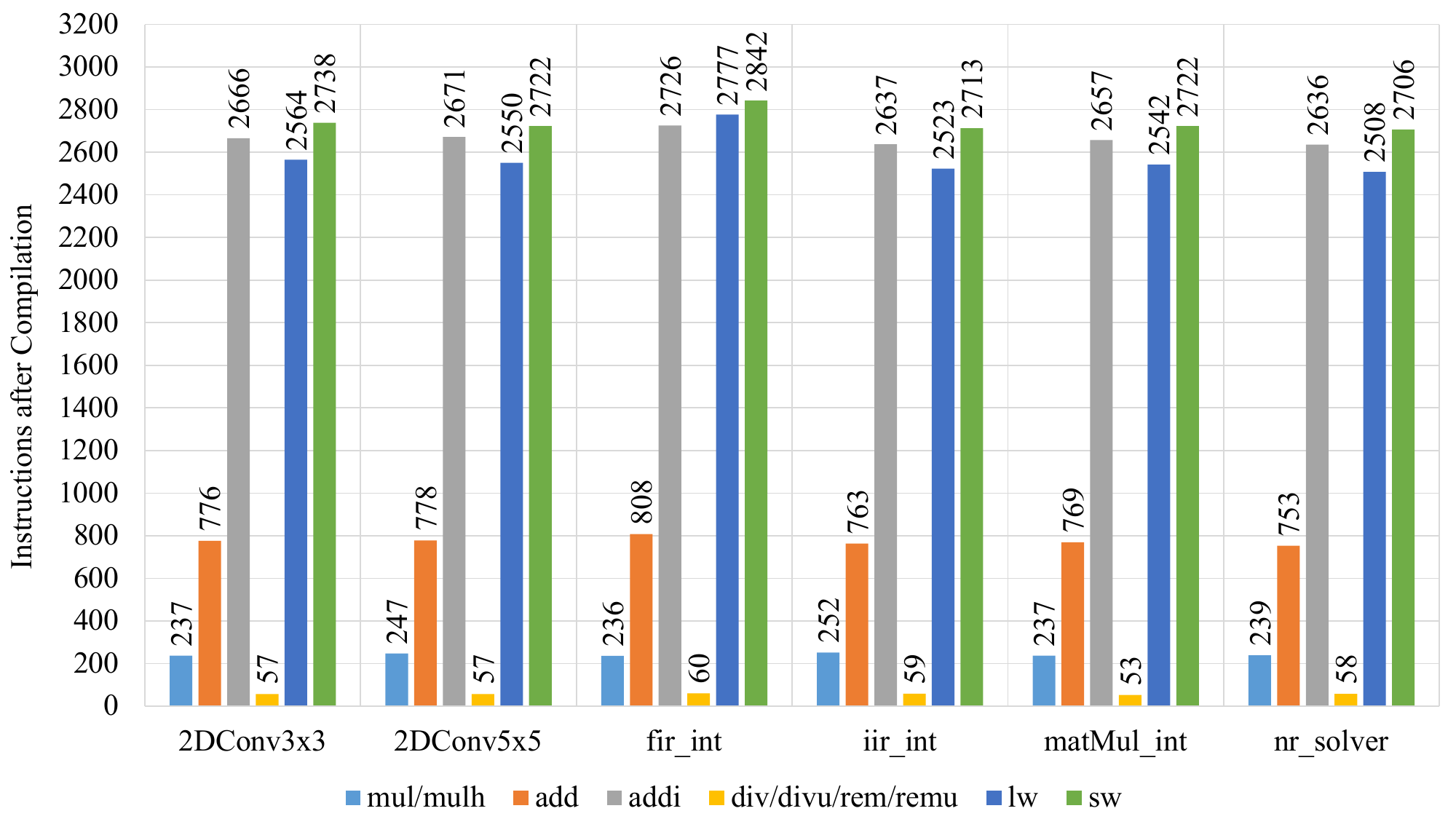}
    \caption{Number of arithmetic and memory accessing instructions after code compilation using GCC compiler \cite{ref-17} with constant options}
    \label{fig-8}
    \end{figure}

    Fig.~\ref{fig-8} presents a detailed breakdown of instruction types following compilation. The average code size is comprised of 221992 instructions and utilization of multiplication is 242 instructions per code in average. Multiplication instructions are 6.47\% of arithmetic instruction (addition/multiplication/division) in average. Approximation is enabled exclusively for multiplication instructions, specifically the \verb|mul| and \verb|mulh| instructions. Both of these instructions are classified as R-type, meaning they operate with registers without involving any immediate values. They utilize the same circuitry, which is determined by the \verb|mulcsr| setting in the program. In contrast, addition instructions (\verb|add| and \verb|addi|) operate in accurate mode.

    There is a selected set of standard applications using a notable load of multiplication within the calculation process. Some of these applications are regarded as fault-tolerant applications due to their inherent resilience against error in accumulative workload. As an example, 2D convolutions are used in image processing applications and filters, which are perfect examples of applications allowed to adhere approximation. All of the programs are compiled using standard GCC-GNU \cite{ref-17} compiler for RISC-V with similar compilation options. Duration for execution of each code and the instruction count is measured using RISC-V standard CSRs in order to calculate energy efficiency of the processor in each application. Energy efficiency in this study is regarded as the consumed power in a limited time period per instructions executed in the program. 

\begin{figure}[htbp]
    \centering
    \includegraphics[width=\linewidth]{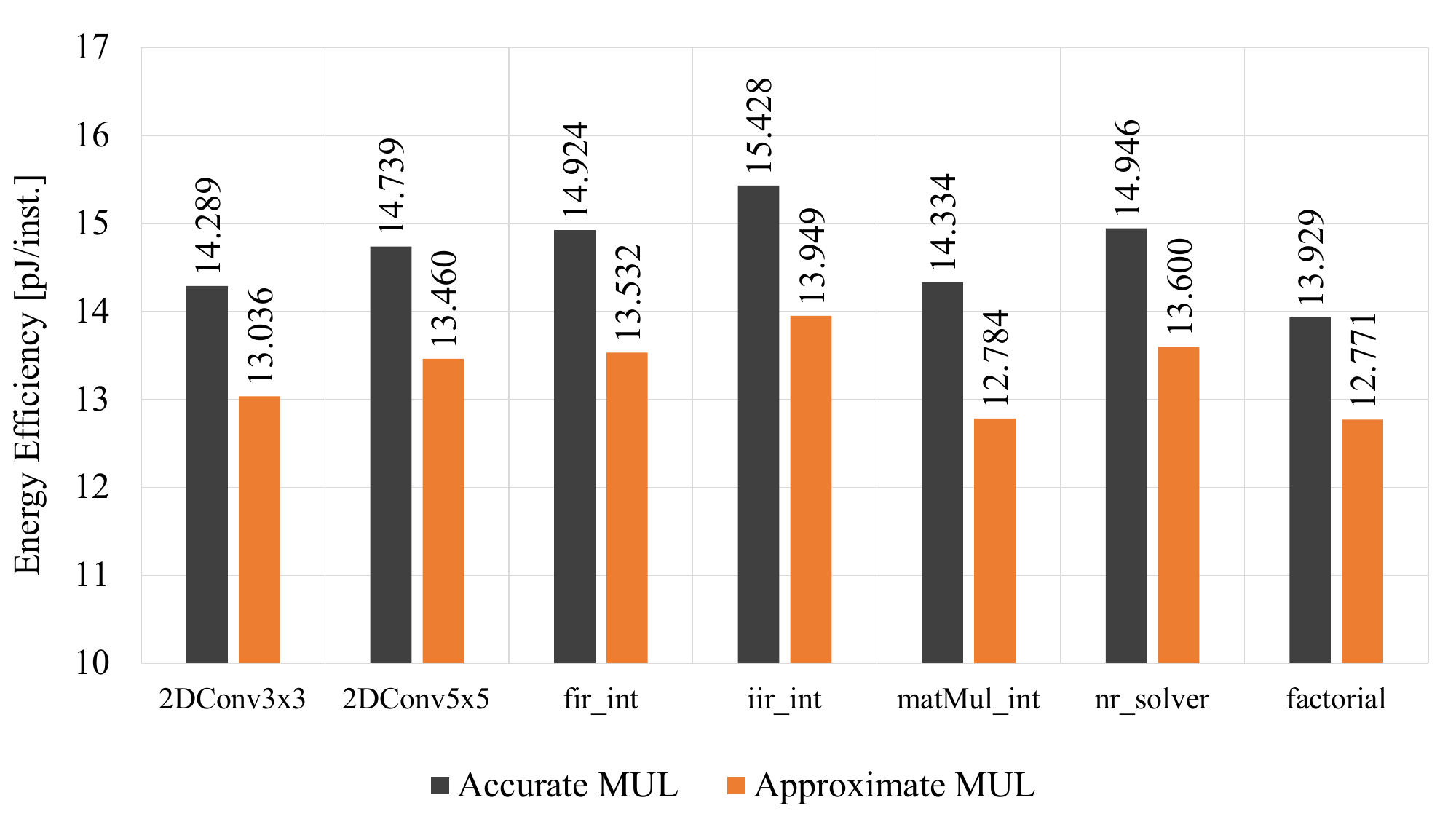}
    \caption{Energy efficiency comparison of applications executed with accurate and approximate multiplication}
    \label{fig-9}
    \end{figure}

\begin{figure}[htbp]
    \centering
    \includegraphics[width=\linewidth]{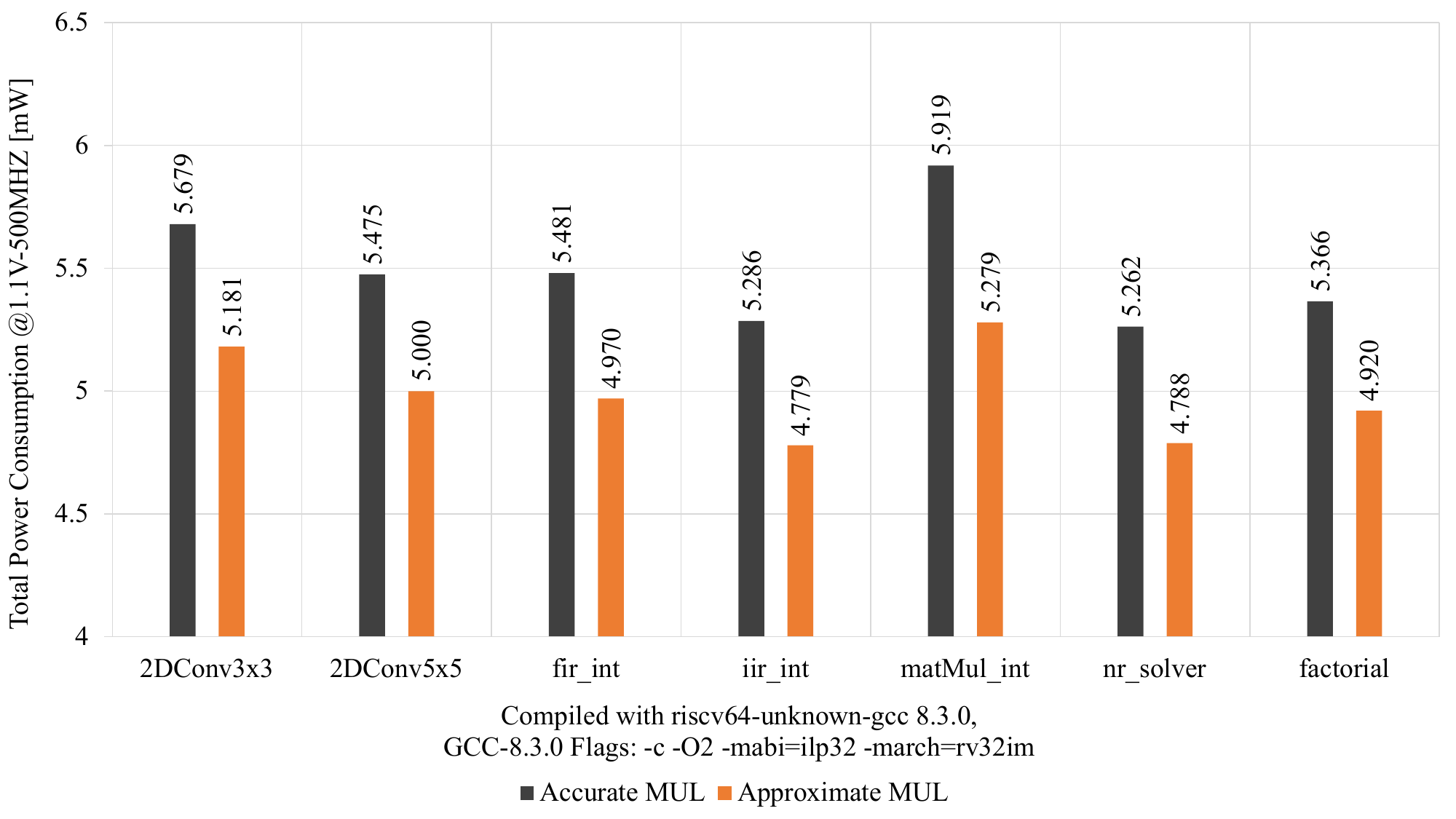}
    \caption {Overall power consumption comparison of the processor, executing applications using accurate and approximate multiplication}
    \label{fig-10}
    \end{figure}

    Fig.~\ref{fig-9} illustrates energy efficiency in each program in pJ/instructions using both accurate and approximate multipliers. Circuit switching was done utilizing the \verb|mulcsr| circuit selection field. The approximate multiplier is set to a default accuracy with error control field of CSR equal to 0x7E.
    Fig.~\ref{fig-10} shows the impact of approximate multiplication on total power consumption of the embedded processor. In every application, the approximate multiplication has decreased overall power consumption. The conserved power in each application highlights the efficiency of approximation techniques in scenarios where a reasonable margin of calculation error is permissible. The selected error-level (0x7E) of the multiplier has an MRED of 0.85\% and ER of 36.2\%.

    Fig.~\ref{fig-10} shows the impact of approximate multiplication on total power consumption of the embedded processor. In every application, the approximate multiplication has decreased overall power consumption. The saved power in each application showcases the effectiveness of approximation in applications where error in calculations are allowed by a reasonable margin. The selected error-level of the multiplier has an MRED of 0.85\% and ER of 36.2\%.

    \begin{figure}[htbp]
    \centering
    \includegraphics[width=\linewidth]{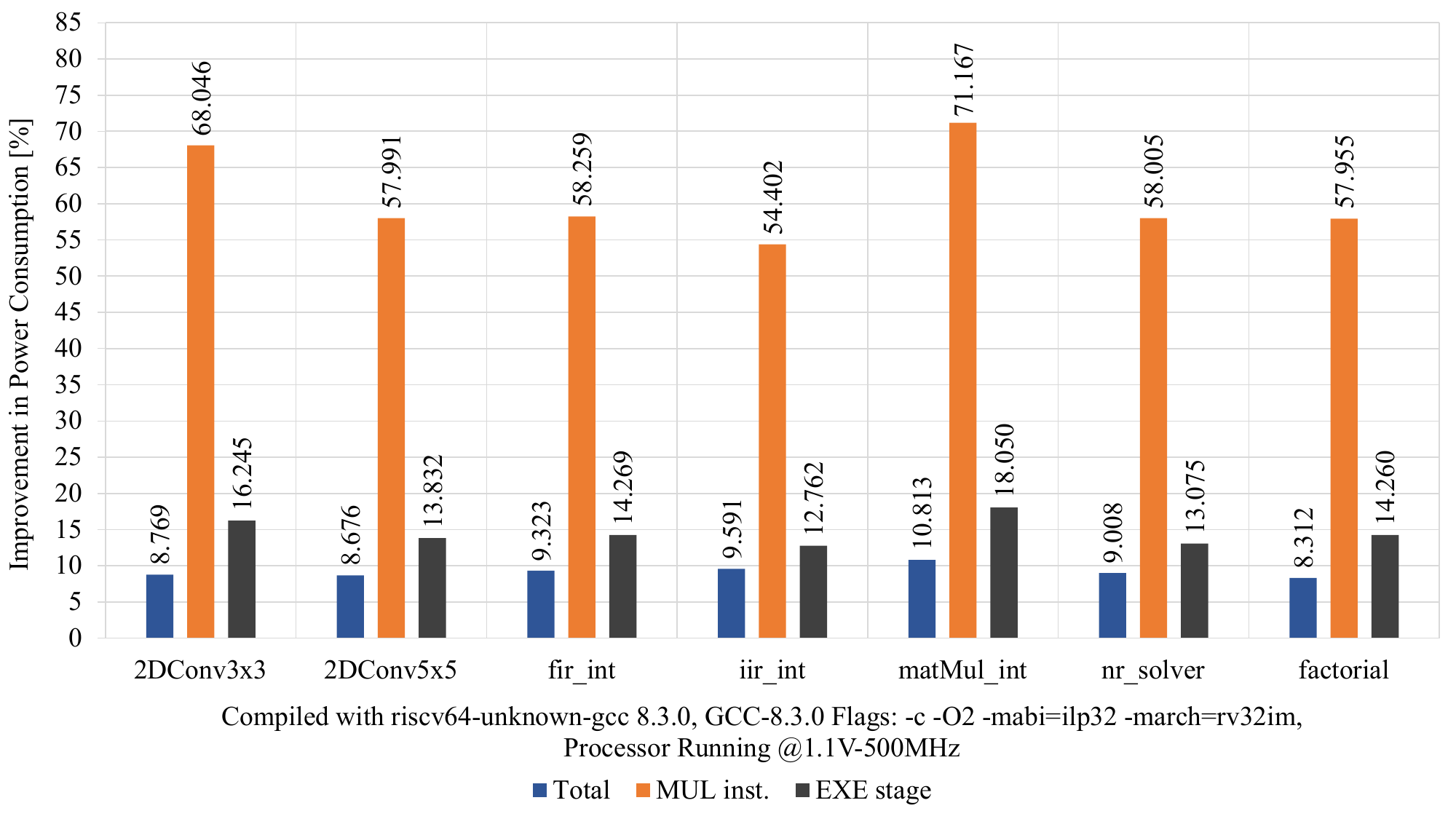}
    \caption {Power consumption improvements by approximation, in the processor's overall power, execution stage and multiplication instructions}
    \label{fig-11}
    \end{figure}

    Overall improvements in term of power consumption are illustrated in Fig.~\ref{fig-11}. This analysis was carried out from 3 separate perspectives: Total improvement (processor’s power consumption), Execution stage of the processor, and the Multiplier Unit (multiplication operation solely). The percentage of improvement in each considered hardware is written next to each column. Power consumption in each execution unit and the execution stage of the core are presented in Table~\ref{table-2} and Table~\ref{table-3} in both accurate and approximate modes. 

    From the results, it is indicated that there is an average improvement of 9.213\% in energy-efficiency in comparison with accurate calculation. By selecting the desired circuit in each execution unit though CSRs, the other circuits will be turned off, decreasing the dynamic and static power consumption of unused units near to zero. The highest improvement is for\textit{matMul\_int} with 10.813\% percent with a power consumption of 5.279mW with approximate multiplier. 
    
    The overall power consumption during the execution stage of the pipelined processor is reduced by 18.05\% through the use of approximate multiplication, with the multiplication process alone achieving a significant improvement of 71.16\% within this stage. The most significant improvement in these metrics is resulted from the\textit{matMul\_int} and \textit{2Dconv3x3}.

        \begin{table}[htbp]
    \caption{Power Consumption of Execution units in Accurate Multiplication Mode}
    \begin{center}

    \noindent\begin{minipage}{\linewidth}
    \centering
    \begin{tblr}{
      width = \linewidth,
      colspec = {Q[100]Q[100]Q[100]Q[100]Q[100]},
      cells = {c},
      cell{1}{1} = {r=2}{},
      cell{1}{2} = {c=4}{0.639\linewidth},
      vlines,
      hline{1,3-10} = {-}{},
      hline{2} = {2-5}{},
    }
    Applications & Power Consumption [mW] &  &  & \\
     & MUL & ALU & DIV & EXE
      Stage\\
    2DConv3x3 & 0.435 & 0.626 & 0.601 & 1.662\\
    2DConv5x5 & 0.438 & 0.627 & 0.605 & 1.670\\
    fir\_int & 0.448 & 0.598 & 0.622 & 1.668\\
    iir\_int & 0.443 & 0.592 & 0.634 & 1.669\\
    matMul\_int & 0.437 & 0.646 & 0.507 & 1.590\\
    nr\_solver & 0.431 & 0.589 & 0.632 & 1.652\\
    factorial & 0.440 & 0.593 & 0.615 & 1.648
    \end{tblr}
    \end{minipage}
    
    \label{table-2}
    \end{center}
    \end{table}


    \begin{table}[htbp]
    \caption{Power Consumption of Execution units in Approximate Multiplication Mode}
    \begin{center}

    \noindent\begin{minipage}{\linewidth}
    \centering
    \begin{tblr}{
      width = \linewidth,
      colspec = {Q[100]Q[100]Q[100]Q[100]Q[100]},
      cells = {c},
      cell{1}{1} = {r=2}{},
      cell{1}{2} = {c=4}{0.639\linewidth},
      vlines,
      hline{1,3-10} = {-}{},
      hline{2} = {2-5}{},
    }
    Applications & Power Consumption [mW] &  &  & \\
     & MUL & ALU & DIV & EXE
      Stage\\
    2DConv3x3 & 0.139 & 0.638 & 0.615 & 1.392\\
    2DConv5x5 & 0.184 & 0.636 & 0.619 & 1.439\\
    fir\_int & 0.187 & 0.607 & 0.636 & 1.43\\
    iir\_int & 0.202 & 0.605 & 0.649 & 1.456\\
    matMul\_int & 0.126 & 0.655 & 0.522 & 1.303\\
    nr\_solver & 0.181 & 0.6 & 0.655 & 1.436\\
    factorial & 0.185 & 0.602 & 0.626 & 1.413
    \end{tblr}
    \end{minipage}
    
    \label{table-3}
    \end{center}
    \end{table}
    
\section{Conclusion}\label{conclusion-section}

    This study examines the impact of dynamic hardware approximation on the energy efficiency of a RISC-V embedded processor with specialized capabilities for approximate computing. The findings demonstrate an average 9.21\% improvement in the processor’s run-time energy efficiency, along with a 14.64\% reduction in overall power consumption during the execute stage of the scalar core. These results confirm that approximate computing, particularly through hardware-approximated arithmetic circuits, is a promising approach for dynamic energy management in embedded processors. The research highlights the potential of such techniques to optimize power usage without compromising computational performance, making them particularly valuable in resource-constrained environments. This study further reinforces the growing relevance of hardware approximation in addressing energy efficiency challenges in modern embedded systems.
    
    The results highlight the potential of integrating RISC-V processors and approximate computing for enhancing the performance of energy-constrained embedded systems in execution process. The ability to dynamically adjust approximation offers a flexible approach to manage accuracy and efficiency. Future work could explore extending these techniques across the processor architecture and to a wider range of embedded applications. Further hardware optimization and advanced error management could lead to even greater efficiency gains. Automating the process of accuracy and energy control using real-time monitoring systems within the processor can be a promising solution to address the accuracy/efficiency trade-off in embedded cores. Overall, this study demonstrates the significant outlook for RISC-V embedded processors leveraging approximate computing.

\section*{Acknowledgment}

The authors would like to express their appreciation to A. M. H. Monazzah for their valuable insights, which have significantly enhanced the quality of this paper.

\end{document}